\documentstyle[buckow]{article}
\def\MM{{\cal M}}
\def\NN{{\cal N}}
\def\WW{{\cal W}}
\def\XX{{\cal X}}
\def\YY{{\cal Y}}

\def\half{\frac12}
\def\d{\partial}

\begin {document}
\begin{flushright}
Bicocca-FT-00-24\\
RI-11-00 \\
hep-th/0012117
\end{flushright}

\def\titleline{
%%%%%%%%%%%%%%%%%%%%%%%%%%%%%%%%%%%%%%%%%%%%%%
%                                            %
% Insert now the text of your title.         %
% Make a linebreak in the title with         %
%                                            %
%            \newtitleline                   %
%                                            %
%%%%%%%%%%%%%%%%%%%%%%%%%%%%%%%%%%%%%%%%%%%%%%
String Theory on $AdS_3$ and Symmetric Products\footnote{Talk given
by R.~Argurio at the workshop ``The quantum structure of
spacetime and the geometric nature of fundamental interactions" in
Berlin, October 2000.}
%\newtitleline
%%%%%%%%%%%%%%%%%%%%%%%%%%%%%%%%%%%%%%%%%%%%%%
}
\def\authors{
%%%%%%%%%%%%%%%%%%%%%%%%%%%%%%%%%%%%%%%%%%%%%%
%                                            %
%  Insert now the name (names) of the author %
%  (authors).                                %
%  In the case of several authors with       %
%  different addresses use e.g.              %
%                                            %
%             \1ad  , \2ad  etc.             %
%                                            %
%  to indicate that a given author has the   %
%  address number 1 , 2 , etc.               %
%                                            %
%%%%%%%%%%%%%%%%%%%%%%%%%%%%%%%%%%%%%%%%%%%%%%
R.~Argurio\1ad, A.~Giveon\2ad and A.~Shomer\2ad
%%%%%%%%%%%%%%%%%%%%%%%%%%%%%%%%%%%%%%%%%%%%%
}
\def\addresses{
%%%%%%%%%%%%%%%%%%%%%%%%%%%%%%%%%%%%%%%%%%%%%%
%                                            %
% List now the address. In the case of       %
% several addresses list them by numbers     %
% using e.g.                                 %
%                                            %
%             \1ad , \2ad   etc.             %
%                                            %
% to numerate address 1 , 2 , etc.           %
%                                            %
%%%%%%%%%%%%%%%%%%%%%%%%%%%%%%%%%%%%%%%%%%%%%%
\1ad
Dipartimento di Fisica\\
Universit\`a Milano Bicocca\\
Piazza delle Scienze 3, 20126 Milano, Italy\\ 
\2ad
Racah Institute of Physics\\
The Hebrew University\\
Jerusalem 91904, Israel
%%%%%%%%%%%%%%%%%%%%%%%%%%%%%%%%%%%%%%%%%%%%%%%
}
\def\abstracttext{
%%%%%%%%%%%%%%%%%%%%%%%%%%%%%%%%%%%%%%%%%%%%%%%
%                                             %
% Insert now the text of your abstract.       %
%                                             %
%%%%%%%%%%%%%%%%%%%%%%%%%%%%%%%%%%%%%%%%%%%%%%%
We consider string theory on a background of the form $AdS_3 \times \NN$.
Our aim is to give a description of the dual CFT in a 
general set up. With the requirement that we have $N=2$ supersymmetry
in spacetime, we provide evidence that the dual CFT is in the moduli space of
a certain symmetric product $\MM^p/S_p$. 
On the way to show this, we reproduce some recent results on
string propagation on $AdS_3$ and extend them to the superstring.
%%%%%%%%%%%%%%%%%%%%%%%%%%%%%%%%%%%%%%%%%%%%%%%%%%%%%%%%%%%%%%%
}
\large
\makefront
%%%%%%%%%%%%%%%%%%%%%%%%%%%%%%%%%%%%%%%%%%%%%%%%
%                                              %
%  Insert now the remaining parts of           %
%  your article.                               %
%                                              %
%%%%%%%%%%%%%%%%%%%%%%%%%%%%%%%%%%%%%%%%%%%%%%%%
% 4 resp. 6 pages!
\section{Introduction}
We consider in this contribution string theory on $AdS_3$ from
the perspective of the $AdS$/CFT correspondence. Namely, 
strings on  $AdS_3$ are dual to a (spacetime) 2-$d$ CFT which
can be thought of as being located on the boundary of the $AdS_3$.
To be more precise, the background on which the strings propagate
is $AdS_3\times \NN$, where $\NN$ is a CFT with suitable central
charge to make the string theory critical. 
The details of the dual spacetime CFT$_2$ will depend on the worldsheet
CFT $\NN$, for instance (super)symmetries of $\NN$
will be reflected in (super)symmetries in spacetime.

It was known since a long time \cite{bh} that the Virasoro algebra
of conformal symmetry of a two-dimensional theory can be
recovered from diffeomorphisms of a theory
of gravity on $AdS_3$. This can of course be extended to 
superconformal symmetry, starting from supergravity on $AdS_3$ \cite{bbcho}.
The central charge of the boundary (super)conformal theory is
already fixed classically, and is given by $c_{st}= {3L\over 2 G_3}$,
where $L$ is the ``radius" of $AdS_3$ (i.e. the cosmological constant
is given by $\Lambda=-{1\over L^2}$)   and $G_3$ is 
the 3-dimensional Newton constant.

It was then realized  in \cite{malda} that the near horizon
geometry of a system of $p$ D1-branes parallel to $k$ D5-branes wrapped,
say, on $T^4$, was given by $AdS_3 \times S^3 \times T^4$. Rewriting
the three dimensional constants $L$ and $G_3$ in terms of the stringy
ones, the spacetime central charge ends up being:
\begin{equation}
c_{st}= 6kp.
\label{stcc}
\end{equation}
Since the system above can be S-dualized to a system where
only NSNS fields are turned on, the possibility opens up to
study the above background perturbatively using string theory
worldsheet techniques \cite{gks,kll}.

To formulate the problem in string perturbation theory, one notices
that $AdS_3$ is nothing else than (the universal covering of) the
group manifold of $SL(2)$.
Thus the worldsheet CFT is taken to be the product of the WZW model on 
$SL(2)$ at level $k$, times another CFT $\NN$.
For some specific such $\NN$, one gets a specific candidate
for the dual CFT$_2$. 
On the other hand, for a general $\NN$ 
one can use the methods of \cite{gr,bl} in order to construct 
the superconformal algebra of the boundary theory.
Our aim is to go beyond this result and
specify some properties that hold in general
for the spacetime CFT. We will indeed find that its chiral spectrum
has a specific structure \cite{ags}.

\section{Review of strings on $AdS_3$}
The study of string theory on $AdS_3$ is an ``old" issue,
whose interest stems from the possibility to handle
perturbatively string propagation on backgrounds including
a curved time (see \cite{early} for a cospicuous list of early
and also more recent references on the subject).

The problems which arose from the beginning were related
to the fact that one is considering a non-compact WZW model.
More precisely, such a model has a host of negative norm
states, and one hopes that in string theory one  will 
get rid of them through imposing the Virasoro constraints
(as it is the case with the timelike oscillators in flat spacetime).
However it turns out in this case that one of the conditions
for the ``no-ghost theorem" on $SL(2)_k$ is the requirement
that the physical states be built only on the (discrete) 
$SL(2)$ representations with spin $j$ in the range $-1<j<{k-2\over 2}$.
Notwithstanding the long-time open problem of the compatibility of
this finite range with modular invariance, 
the upper bound is also particularly counter-intuitive because of the
following. When considering string theory on $AdS_3 \times \NN$,
the bound $j<{k-2\over 2}$ implies, through the on-shell condition,
that for physical states there is a bound on the weight $\Delta$
of operators in $\NN$. For instance, if there is a cycle $S^1 \subset \NN$,
then the winding number would be bounded by virtue of the upper bound
on $j$.

Together with this older problem, a new one arose in the context
of the $AdS$/CFT correspondence, which revived interest in string
theory on $AdS_3$. It is the problem of the perturbative formulation
of the so-called `long strings' \cite{mms,sw,ks}, related to the number $p$. 
The presence of these long strings was already postulated in the
work of \cite{gks}, in particular in the way the spacetime central
charge (\ref{stcc}) was derived.
Another crucial remark was the one of \cite{mms} that a BPS string
in an $AdS_3$ background could nucleate at the origin and expand
to the boundary (where it has an infinite spatial extent) at a
finite cost in energy.
This analysis was refined in \cite{sw} where it was shown that
the possibility to produce long strings should be seen in the
boundary CFT by the presence of a continuum of states (roughly, the
radial momentum of the long string) above a gap (the cost in energy
to pull a long string to the boundary).

From the above we learn that, if the worldsheet theory gives a good
description of the spacetime CFT, we should expect to find there
vertex operators corresponding in spacetime to this continuum of states.
These vertex operators were not seen in the `traditional' $SL(2)$
WZW approach.

However, Maldacena and Ooguri offered in \cite{mo}
a new look on the $SL(2)$ WZW model
that proposes a solution to both of the above problems.
Their main argument is that the Hilbert space of the $\widehat{SL(2)}$
WZW model must include a host of new (affine) representations,
obtained from the usual ones by spectral flow. 
This was first noticed in the work of Hwang and collaborators
in Ref.~\cite{early}.
Most of the new representations feature an $L_0$ which is 
unbounded from below. This is however not
a problem since in a string theory context one has to impose
the Virasoro constraints.

Let us sketch very briefly their reasoning, starting by defining the 
spectral flow transformation:
\begin{equation}
\begin{array}{ccl}
J^\pm_n & \rightarrow & J^\pm_{n\pm w}\\
J^3_n  & \rightarrow & J^3_n -{k\over 2} w \delta_n \qquad \qquad w 
\in {\bf Z}\\
L_n  & \rightarrow & L_n - w J^3_n  -{k\over 4} w^2 \delta_n.
\end{array}
\label{spfl}
\end{equation}
By applying the above transformations on the various representations
of $\widehat{SL(2)}$, one obtains the following. The spectral flow of
the discrete representations $\hat{\cal D}^\pm_j$
(which are defined by having an affine
primary field with $SL(2)$ spin $j$ real and bounded by 
$-\half<j<{k-3\over 2}$ \cite{gk,mo}, and a momentum $m$  given
by $m=\pm(j+n)$,
with $n$ a positive integer) gives rise to vertex operators whose
$SL(2)$ part has an increasingly negative $L_0$, and thus allows 
the weight $\Delta$ of the part relating to $\NN$ to grow without
bounds. 
On the other hand, the spectral flow of the continuous representations
$\hat{\cal C}_{j,\mu}$
(which are defined by an $\widehat{SL(2)}$ primary with 
$j=-\half + is$ and $m=\mu,\mu\pm n$, with 
$0\leq \mu <1$) corresponds to the continuous spectrum of the long strings.
Note that the non-spectrally flown ($w=0$)
continuous representation is usually associated with the tachyon,
and is actually projected out in the superstring, as we discuss later.
Thus here the spectral flow is essential to recover the signature
of the long strings in the worldsheet theory.
Let us conclude this section by noting that in the work of \cite{mo,mos}
a modular invariant partition function of the (bosonic) $SL(2)$ WZW
model is explicitely worked out.

\section{Strings on $AdS_3$ and a twist field}
Let us now address the same issues as above but in a different
formulation that will allow us to extend the results to the
superstring and compute the spacetime chiral
spectrum in a convenient way \cite{ags}.

Starting from the bosonic WZW model $SL(2)_k$, the holomorphic currents
satisfy the following OPE (the same holds for the anti-holomorphic
ones):
\begin{equation}
J^A (z) J^B(0) \sim {{k\over 2} \eta^{AB} \over z^2 }
+{i\epsilon^{ABC}\eta_{CD}J^D(0) \over z}, 
\label{opej}
\end{equation}
where $A=1,2,3$ and $\eta_{AB}=(+,+,-)$.
Rewriting:
\begin{equation}
J^3=-\sqrt{k\over 2} \d X,
\label{j3}
\end{equation}
a primary field of the $SL(2)$ WZW model can be written as:
\begin{equation}
\Phi_{jm\bar{m}}=\Psi_{jm\bar{m}}
e^{\sqrt{2\over k}(mX(z)+\bar m\bar X(\bar z))}~,
\label{phipsi}
\end{equation}
where $\Psi_{jm\bar{m}}$ can be thought of as an $SL(2)/U(1)$
parafermion, and the weight of the primary field is:
\begin{equation}
\Delta(\Phi_{jm\bar{m}})={-j(j+1)\over k-2}.
\label{dimphi}
\end{equation}
The primary fields $\Phi_{jm\bar{m}}$ belong to one of the following
representations of $SL(2)$: lowest or highest weight discrete representations
${\cal D}^\pm_j$, or continuous representations ${\cal C}_{j,\mu}$.

An important constraint which must be satisfied by the operators
$\Phi_{jm\bar{m}}$ for single valuedness of the wave function on
$AdS_3$ is that:
\begin{equation}
m-\bar{m}\in {\bf Z}~.
\label{constr}
\end{equation}
In \cite{gks,ks} the above condition was introduced by hand. Here we
wish to impose it more intrinsically, and a way to do that is to
introduce ``twist fields" which implement (\ref{constr}) by
requiring mutual locality of the worldsheet operators  with them.
Let us stress that we are not twisting an otherwise consistent theory,
rather it is consistency of the theory
(tree level unitarity and higher loop modular invariance)
which requires the inclusion
of these twist fields, and of all the twisted operators which arise through
OPEs with the twist fields.

We thus introduce the operators:
\begin{equation}
t^w=e^{w\sqrt{k\over 2}(X(z)+\bar{X}(\bar{z}))}~, \qquad w\in {\bf Z}~.
\label{twist}
\end{equation}
Then the OPE of $t^w$ with $\Phi_{jm\bar{m}}$ takes the form
\begin{equation}
\begin{array}{cl}
t^w(z,\bar z)\Phi_{jm\bar{m}}(z',\bar z')&\sim
(z-z')^{-wm}(\bar z-\bar z')^{-w\bar m}\Phi_{jm\bar{m}}^w(z',\bar z')\\
&=(z-z')^{-(m-\bar m)w}|z-z'|^{-2\bar m w}\Phi_{jm\bar{m}}^w~,
\end{array} \label{mutloc}
\end{equation}
where
\begin{equation}
\Phi_{jm\bar{m}}^w=\Psi_{jm\bar{m}}
e^{\sqrt{2\over k}[(m+{k\over 2}w)X+(\bar m+{k\over 2}w)\bar X]}~.
\label{phitw}
\end{equation}
{}From (\ref{mutloc}) we see that mutual locality with $t^w$ indeed
implies the condition (\ref{constr}). Now, consistency of the theory
implies that we must include also the ``twisted'' operators
$\Phi_{jm\bar{m}}^w$. 

The scaling dimension of $\Phi_{jm\bar{m}}^w$ is:
\begin{equation}
\Delta(\Phi_{jm\bar{m}}^w)=
{-j(j+1)\over k-2}-{k\over 4}w^2-mw~.
\label{dimphitw}
\end{equation}
Then for a vertex operator of $AdS_3 \times \NN$ in the $w$ twisted sector:
\begin{equation}
V^w=V_\Delta \Phi_{jm\bar{m}}^w~,
\label{physop}
\end{equation}
where $V_\Delta$ is a primary
operator in the CFT on $\NN$ with weight $\Delta$,
the on-shell condition is:
\begin{equation}
-{j(j+1)\over k-2}-{k\over 4}w^2-mw+\Delta=1~.
\label{physcon}
\end{equation}
In spacetime it creates from the vacuum a state with weight given by:
\begin{equation}
h=|m| + {k\over 2} |w|, 
\label{weigen}
\end{equation}
where for convenience (see \cite{ags} for the details) we choose
fields belonging to the ${\cal D}^-_j$ representation and $w<0$.
(Roughly, recall that the eigenvalue of the spacetime $L_0^{st}$ is
given by minus the eigenvalue of $J^3$ on the worldsheet \cite{gks}.)
When putting together (\ref{physcon}) and (\ref{weigen}),
we get the following expression for the spacetime weight (for $w\neq 0$):
\begin{equation}
h={k|w|\over 4}+{1\over |w|}\left(-{j(j+1)\over k-2}+\Delta-1
\right)~,
\label{weimo}
\end{equation}
in agreement with \cite{mo}. 
In the untwisted sector $w=0$ we had of course $h=j+1$ \cite{gks}.

\subsection*{The superstrings}
Going now to the superstring (we use the NSR
formulation), we have to consider the supersymmetric $SL(2)$
WZW model. In addition to the three currents $J^A$ (we will now
concentrate for simplicity on holomorphic fields)
we have their three fermionic partners $\psi^A$ with OPE:
\begin{equation}
\psi^A(z)\psi^B(0) \sim {\eta^{AB} \over z}.
\label{opepsi}
\end{equation}
Accordingly, the currents decompose in bosonic and fermionic
pieces:
\begin{equation}
J^A=j^A - {i\over 2} {\epsilon^{A}}_{BC} \psi^B \psi^C,
\label{totcurr}
\end{equation}
where $j^A$ has a regular OPE with the fermions. Taking the total
current to be at level $k$, it follows that the bosonic part has
level $k+2$ since the fermionic piece has level $-2$. Note that
this entails a slight change in the value of certain quantities.
For instance, the weight of a bosonic primary of the WZW model is
now $\Delta(\Phi_{jm})=-{j(j+1)\over k}$, and the range for the
$SL(2)$ spin of the discrete representations ${\cal D}^\pm_j$ is:
\begin{equation}
-\half < j < {k-1\over 2}.
\label{bound}
\end{equation}
In the following, in order to have some robust information
on the spectrum of the spacetime theory, we will need to have 
at least $N=2$ supersymmetries in spacetime. The conditions to achieve
that were studied in \cite{gr,bl}. There it was shown that 
string theory on $AdS_3 \times \NN$ gives rise to a boundary $N=2$ SCFT
provided that we can write, at least locally, $\NN=U(1)_Y \times
\NN / U(1)$, and that $\NN/U(1)$ is a worldsheet CFT with $N=2$
supersymmetry (and central charge $c_{\NN/U(1)}=9-{6\over k}$). 
For later convenience, we write the current corresponding to the
$U(1)_Y$ in terms of the canonically normalized scalar $Y$:
\begin{equation}
J^Y=i\d Y.
\label{u1curr}
\end{equation}
With the above conditions, 
the {\it spacetime} supercharges $G^\pm_r$ can be constructed \cite{gr,bl}, 
together with all the superconformal generators.

Let us single out the following piece of the algebra generated by the 
supercharges:
\begin{equation}
\{G^+_r,G^-_s\}  =2L_{r+s}+(r-s)J_0 ~,\qquad r,s=\pm\frac12,
\label{supalg}
\end{equation}
where the zero mode $J_0$ of the spacetime $R$-current is given
by the worldsheet operator:
\begin{equation}
J_0=\sqrt{2k}\oint dz J^Y(z)~.
\label{strcurr}
\end{equation}
The twist field in the supersymmetric case has to be slightly
amended, in order for it to be mutually local with the
supercharges, and also convenient for the study of the
chiral spectrum.
We thus introduce:
\begin{equation}
t^w_\pm = e^{-w\int J^3 \pm \frac{w}2\int J^Y}=e^{w\sqrt{\frac{k}2}(X\pm iY)}.
\label{susytw}
\end{equation}
Note that the $J^3$ above is the total current. The twist fields $t^w_\pm$,
being mutually local with $G^\pm_r$, survive the GSO projection
(thus the twist of a GSO invariant operator is also GSO invariant),
have zero weight on the worldsheet, and are ``chiral" in spacetime
in the sense that they verify $|J^3|=\half |J^Y|$.
 
\section{Chiral spectrum, untwisted and twisted}
We now give a worldsheet description of operators which are chiral
(or anti-chiral) under the spacetime superconformal algebra (\ref{supalg}). 
Let us first of all concentrate on vertex operators in the
untwisted sector $w=0$. Moreover, we consider for now only operators whose
$SL(2)$ piece is in discrete representations ${\cal D}^\pm_j$.

Every such vertex operator will include, besides a piece relating to the
bosonized superghost, pieces relating to the $U(1)_Y$,
the $\NN/U(1)$ and the $SL(2)$ factors of the worldsheet  CFT.
The on-shell and the GSO conditions together tie the respective values
of the momentum along the $U(1)_Y$, the weight and $R$-charge in
$\NN/U(1)$ and the $SL(2)$ spin $j$. 
After imposing all these conditions in both the NS and R sectors, we obtain
the following physical vertex operator which are (anti)chiral in spacetime.

In the NS sector, there are two kinds of vertex operators. The first family
is given by:
\begin{equation} 
\XX=e^{-\varphi}e^{-i\sqrt{2\over k}(j+1)Y}V
\Phi_{j={k\over 2}(1-r_V)-1,m}~,\qquad\qquad \Delta_V={r_V\over 2}, \qquad
r_V+{1\over k}<1~,
\label{chix}
\end{equation}
where the inequality is implied by (\ref{bound}). The spacetime weight
and $R$-charge of these operators are:
\begin{equation}
h_\XX={k\over 2}(1-r_V)=-\half R_\XX.
\label{weix}
\end{equation}
The operators $\XX$ are thus antichiral. Note that the corresponding 
chiral operators (with positive spacetime $R$-charge) are simply
obtained by charge conjugation (and will involve $V$ with $r_V<0$;
here and below we restrict for simplicity to $r_V\geq 0$).

The second family of vertex operators is:
\begin{equation}
\WW=e^{-\varphi}e^{i\sqrt{2\over k}jY}V
(\psi\Phi_{j={k\over 2}r_V})_{j-1,m}~,\qquad\qquad \Delta_V={r_V\over 2}, 
\qquad r_V+{1\over k}<1.
\label{chiw}
\end{equation}
Their spacetime weight is:
\begin{equation}
h_\WW={k\over 2}r_V=\half R_\WW.
\label{weiw}
\end{equation}
In the Ramond sector we only have one family:
\begin{equation}
\YY=e^{-{\varphi\over 2}}(S e^{i\sqrt{2\over k}jY}
V \Phi_{j={k\over 2}(r_V-1)})_{j-\half,m}, \qquad \qquad \Delta_V={r_V\over 2},
\qquad 1<r_V+{1\over k}<2~.
\label{chiy}
\end{equation}
Here $S$ is a spinfield whose precise expression can be found in \cite{ags}.
The spacetime weight of the $\YY$ vertex operators is:
\begin{equation}
h_\YY=\half + {k\over 2}(r_V-1)=\half R_\YY.
\label{weiy}
\end{equation}
Note that all the above vertex operators give rise to spacetime (anti-)chiral
operators whose weight lies in the range $0\leq h \leq {k\over 2}$. 

Going now to the twisted sectors, $w\neq 0$, the task of finding the
new operators which are chiral in spacetime is simplified by our
choice of twist fields (\ref{susytw}). It turns out that it is 
sufficient to twist the above physical operators (\ref{chix}),
(\ref{chiw}) and (\ref{chiy}). The physicality condition on the
twisted operators will only fix the value of $m$ (this essentially
means that we obtain in spacetime only the Virasoro highest weight).
The physical vertex operators that we get in both the NS and R sector
are thus:
\begin{equation}
\begin{array}{cclcl}
\XX^w&=&e^{-\varphi}e^{-i\sqrt{2\over k}(j+1-{kw\over 2})Y}V \Phi^w_{j
={k\over 2}(1-r_V)-1,m=-j-1}~, &\quad& h_\XX^w=
{k\over 2}(1-r_V)+{k|w|\over 2}, \\
\WW^w&=&e^{-\varphi}e^{i\sqrt{2\over k}(j-{kw\over 2})Y}V
(\psi\Phi^w_{j={k\over 2}r_V})_{j-1,m=-j}~,&\quad& h_\WW^w=
{k\over 2}r_V+{k|w|\over 2},\\
\YY^w&=& e^{-{\varphi\over 2}}(Se^{i\sqrt{2\over k}(j-{kw\over 2})Y}
V \Phi^w_{j={k\over 2}(r_V-1)})_{j-\half,m=-j-\half}~, &\quad&
h_\YY^w= \half +{k\over 2}(r_V-1)+{k|w|\over 2} .
\end{array}
\label{chitw}
\end{equation}
A clear pattern emerges: every operator in the untwisted sector
gives rise to a tower of operators, regularly spaced with a step
of $k/2$, which are all (anti-)chiral in spacetime. We thus have
a spacetime theory whose chiral spectrum is ordered according to
the following pattern:
\begin{equation}
h^w=h^0+{k\over 2}|w|. 
\label{pattern}
\end{equation}
We should now briefly comment on the vertex operators which belong
to the continuous representations ${\cal C}_{j,\mu}$ of $SL(2)$.
In the untwisted sector, they are tachyonic and do not survive the GSO
projection. However, in the twisted sectors such operators can be
physical, and give rise to the long string continuous spectrum, as
in \cite{mo}. Moreover, we expect also that some chiral states
can be found within the continuous spectrum, but to find and classify them
is a more complicated issue since one cannot simply twist already
known physical states, as with the discrete spectrum.

\section{The spacetime CFT as a symmetric product}
We will now provide evidence that the spacetime CFT is a deformation
of the following symmetric product CFT:
\begin{equation}
(\MM_{c=6k})^p/S_p,
\label{sympro}
\end{equation}
which indeed has a central charge $c_{st}=6kp$. We will specify what
$\MM_{6k}$ is after showing that the pattern (\ref{pattern}) is
reproduced in the chiral spectrum of a CFT like (\ref{sympro}).
For earlier studies of symmetric product CFTs, see \cite{spref}.

To be more precise, we focus on the chiral spectrum of ``single 
particle states," since this is what the string spectrum can 
be compared to: vertex operators correspond to single string
states in spacetime. In a CFT like (\ref{sympro}), the
``single particle states" are identified with the ones
coming from each single $Z_N$ twisted sector of the $S_p$ orbifold. Note
that this includes also the ``diagonal" $\MM_{6k}$ as the $Z_1$ sector.
In order to obtain the spectrum of the $Z_N$ twisted sector, we consider
the $\MM^N/Z_N$ orbifold. There is a one-to-one correspondence between
states in $\MM$ with weight $h$ and $R$-charge $R$ 
and states in the $Z_N$ twisted sector of $\MM^N/Z_N$ with \cite{klsc}:
\begin{equation}
h^N={h\over N} + {c\over 24} {N^2-1\over N}, \qquad \qquad R^N=R.
\label{zn}
\end{equation}
Another useful property of $N=2$ SCFTs is the existence of a spectral
flow between the R and the NS sectors \cite{ss}:
\begin{equation}
h_R=h_{NS}-\half R_{NS} + {c\over 24}, \qquad \qquad
R_R=R_{NS} -  {c\over 6}.
\label{rnsflo}
\end{equation}
We can now proceed to show that every chiral operator of $\MM$ gives
rise to a chiral operator in the $Z_N$ twisted sector of $\MM^N/Z_N$,
and thus of each $Z_N$ twisted sector of $\MM^p/S_p$.

Starting from a chiral state of $\MM$ with $R$-charge $R$ and weight
$h=R/2$, the spectral flow (\ref{rnsflo}) 
to the R sector of $\MM$ gives a state
with $h_R={c\over 24}$ and $R_R=R-{c\over 6}$, which is a Ramond ground state.
Going then to the R sector of $\MM^N/Z_N$ using (\ref{zn}), we get a state with 
$h^N_R={Nc\over 24}$ and $R^N_R=R-{c\over 6}$, which is still a
Ramond ground state, but of the $Z_N$ orbifold CFT. Eventually, we go to
the NS sector of $\MM^N/Z_N$ using (\ref{rnsflo}) backwards, and get
a state with $h^N_{NS}={R\over 2}+{c\over 12}(N-1)$ and $R^N_{NS}=
R+{c\over 6}(N-1)$, which is thus chiral. For $c=6k$, we get the following
pattern:
\begin{equation}
h^N=h^1+{k\over 2} (N-1).
\label{pattern2}
\end{equation}
It is exactly the same pattern as (\ref{pattern}) if we identify $h^{w=0}$
and $|w|$ there with $h^{N=1}$ and $N-1$ in (\ref{pattern2}).
Namely, we take $\MM_{c=6k}$ such that its chiral spectrum is the
one computed from string vertex operators coming from the untwisted
sector $w=0$. 

Note that here we have a precise bound on $N$, namely the largest cycle
is $Z_p$ and thus we have $N\leq p$. On the other hand, a bound on $w$
is not seen in the worldsheet analysis because it should be non-perturbative
in nature, $p$ being related to the string coupling constant by
$g_s^2\sim 1/p$.

\section{Conclusions and Discussion}
In this contribution, we have provided evidence that the single particle
chiral spectrum of superstring theory on $AdS_3 \times \NN$
and of the symmetric product CFT $(\MM_{6k})^p/S_p$ agree, where
we take $\MM_{6k}$ to have the chiral spectrum provided by
the $w=0$ sector of the worldsheet theory.
More precisely, the $Z_N$ twisted sectors of $(\MM_{6k})^p/S_p$
contains (single string) states associated with the worldsheet
theory sector related to the presence of $N$ long strings.

It is likely that the same pattern exists (albeit deformed
by some amount) for the non chiral spectrum, and for less
or non supersymmetric string theories.

Let us end with a particular example in which we can be more specific,
and which reveals more structure. Consider string theory on
$AdS_3 \times S^3 \times T^4$ \cite{gks} which has (small) $N=4$
superconformal symmetry. Here the dual CFT is supposed to be a deformation of
$(T^4)^{kp}/S_{kp}$ (plus something that is not seen on the 
worldsheet, see \cite{mmoores}). Our conjecture being (\ref{sympro}),
we would like to postulate that $(T^4)^{kp}/S_{kp}$ is related in some
way to the ``double" symmetric product $((T^4)^k/S_k)^p/S_p$, thus
identifying $\MM_{6k}$ with a theory in the moduli space of
$(T^4)^k/S_k$. Given the conjecture
(\ref{sympro}), the above argument is compatible with U-duality of the brane
configuration leading to that particular background.
Note however that (as it was remarked in \cite{sw}) there are 
missing chiral states in the worldsheet description. This was argued 
\cite{sw} to be
traced to the fact that the CFT dual to the stringy description
is actually sitting on a singular point of the moduli space (roughly
because all the RR fields are vanishing in this background). A
manifestation of the singularity is the presence of the continuum
of the long strings).

This phenomenon where the theory $\MM_{6k}$ could itself be in the moduli
space of a symmetric product, is however not general. Counter examples
can be found in less supersymmetric cases, for instance the pure $N=2$ case
of \cite{gkp} or the $N=3$ case of \cite{ags3}, where it is
shown that $\MM_{6k}$ cannot be a symmetric product of a smaller CFT.

\vskip0.5cm
\noindent
{\large \bf Acknowledgements}

\smallskip
\noindent
We are grateful to D.\ Kutasov for many valuable discussions and
useful comments.
This work is supported in part 
by the European Commision TMR program HPRN-CT-2000-00131, wherein
the University of Milano-Bicocca is associated to the University of Padova,
and in part by the BSF -- American-Israel Bi-National
Science Foundation -- and by the Israel
Academy of Sciences and Humanities -- Centers of Excellence Program.

%%%%%%%%%%%%%%%%%%%%%%
%%%%%%%%%%%%%%%%%%%%%%


\begin{thebibliography}{77}


\bibitem{bh} J.~D.~Brown and M.~Henneaux,
Commun.\ Math.\ Phys.\  {\bf 104} (1986) 207.

\bibitem{bbcho} M.~Banados, K.~Bautier, O.~Coussaert, M.~Henneaux and M.~Ortiz,
Phys.\ Rev.\  {\bf D58} (1998) 085020, hep-th/9805165.

\bibitem{malda} J.~Maldacena,
Adv.\ Theor.\ Math.\ Phys.\  {\bf 2} (1998) 231,
hep-th/9711200;
J.~Maldacena and A.~Strominger, JHEP {\bf 9812} (1998) 005,
hep-th/9804085;
O.~Aharony, S.~S.~Gubser, J.~Maldacena, H.~Ooguri and Y.~Oz,
Phys.\ Rept.\  {\bf 323} (2000) 183,
hep-th/9905111.

\bibitem{gks} A.~Giveon, D.~Kutasov and N.~Seiberg,
Adv.\ Theor.\ Math.\ Phys.\  {\bf 2} (1998) 733,
hep-th/9806194.

\bibitem{kll} D.~Kutasov, F.~Larsen and R.~G.~Leigh,
Nucl.\ Phys.\  {\bf B550} (1999) 183, hep-th/9812027.

\bibitem{gr} A.~Giveon and M.~Rocek, JHEP {\bf 9904} (1999) 019,
hep-th/9904024.

\bibitem{bl} D.~Berenstein and R.~G.~Leigh,
Phys.\ Lett.\  {\bf B458} (1999) 297,
hep-th/9904040.

\bibitem{ags} R.~Argurio, A.~Giveon and A.~Shomer, 
JHEP {\bf 0012} (2000) 003, hep-th/0009242.

\bibitem{early}
A. B. Zamolodchikov and V. A. Fateev, Sov. J. Nucl. Phys. {\bf 43}
(1986) 657;
J. Balog, L. O'Raifeartaigh, P. Forgacs, and A. Wipf,
Nucl. Phys. {\bf B325} (1989) 225;
L. J. Dixon, M. E. Peskin and J. Lykken,  Nucl. Phys. {\bf B325} (1989)
329;
A. Alekseev and S. Shatashvili, Nucl. Phys. {\bf B323} (1989) 719;
N. Mohameddi, Int. J. Mod. Phys. {\bf A5} (1990) 3201;
P. M. S. Petropoulos, Phys. Lett. {\bf B236} (1990) 151;
S. Hwang, Nucl. Phys. {\bf B354} (1991) 100;
M. Henningson and S. Hwang, Phys. Lett. {\bf B258} (1991) 341;
M. Henningson, S. Hwang, P. Roberts, and B. Sundborg, Phys. Lett. {\bf
B267} (1991) 350;
S. Hwang, Phys. Lett. {\bf B276} (1992) 451, hep-th/9110039;
S.~Hwang and P.~Roberts, hep-th/9211075;
K. Gawedzki, hep-th/9110076;
I. Bars and D. Nemeschansky, Nucl. Phys. {\bf B348} (1991) 89;
I. Bars, Phys. Rev. {\bf D53} (1996) 3308, hep-th/9503205;
in {\it Future Perspectives In String Theory} (Los Angeles, 1995),
hep-th/9511187;
Y. Satoh, Nucl. Phys. {\bf B513} (1998) 213, hep-th/9705208;
J. Teschner, Nucl.Phys. {\bf B546} (1999) 390, hep-th/9712256,
Nucl.Phys. {\bf B546} (1999) 369, hep-th/9712258,
Nucl.Phys. {\bf B571} (2000) 555, hep-th/9906215;
J.~M.~Evans, M.~R.~Gaberdiel and M.~J.~Perry,
Nucl.\ Phys.\  {\bf B535} (1998) 152, hep-th/9806024;
J.~de Boer, H.~Ooguri, H.~Robins and J.~Tannenhauser,
JHEP {\bf 9812} (1998) 026, hep-th/9812046;
K.~Hosomichi and Y.~Sugawara, JHEP {\bf 9901} (1999) 013, hep-th/9812100,
JHEP {\bf 9907} (1999) 027, hep-th/9905004;
M.~Yu and B.~Zhang, Nucl.\ Phys.\  {\bf B551} (1999) 425, hep-th/9812216;
N.~Berkovits, C.~Vafa and E.~Witten, JHEP {\bf 9903} (1999) 018,
hep-th/9902098;
I.~Bars, C.~Deliduman and D.~Minic, hep-th/9907087;
P.~M.~Petropoulos, hep-th/9908189;
G.~Giribet and C.~Nunez, JHEP {\bf 9911} (1999) 031, hep-th/9909149;
A.~Kato and Y.~Satoh, Phys.Lett. {\bf B486} (2000) 306, hep-th/0001063;
N.~Ishibashi, K.~Okuyama and Y.~Satoh, Nucl.Phys. {\bf B588} (2000) 149,
hep-th/0005152;
Y. Hikida, K. Hosomichi and Y. Sugawara, Nucl.\ Phys.\ {\bf B589} (2000) 134,
hep-th/0005065;
G.~Giribet and C.~Nunez, JHEP{\bf 0006} (2000) 033, hep-th/0006070.

\bibitem{mms} J. Maldacena, J. Michelson and A. Strominger,
JHEP {\bf 9902} (1999) 011, hep-th/9812073.

\bibitem{sw} N.~Seiberg and E.~Witten,
JHEP {\bf 9904} (1999) 017,
hep-th/9903224.

\bibitem{ks} D.~Kutasov and N.~Seiberg, JHEP {\bf 9904} (1999) 008,
hep-th/9903219.

\bibitem{mo} J.~Maldacena and H.~Ooguri, hep-th/0001053.

\bibitem{gk} A. Giveon and D. Kutasov, JHEP {\bf 9910} (1999) 034, 
hep-th/9909110; JHEP {\bf 0001} (2000) 023, hep-th/9911039.

\bibitem{mos} J.~Maldacena, H.~Ooguri and J.~Son, hep-th/0005183.

\bibitem{spref} C.~Vafa and E.~Witten, Nucl.\ Phys.\  {\bf B431} (1994) 3,
hep-th/9408074; R.~Dijkgraaf, G.~Moore, E.~Verlinde and H.~Verlinde,
Commun.\ Math.\ Phys.\  {\bf 185} (1997) 197,
hep-th/9608096; J.~de Boer, Nucl.\ Phys.\  {\bf B548} (1999) 139,
hep-th/9806104; P.~Bantay, Phys.\ Lett.\  {\bf B419} (1998) 175
hep-th/9708120, hep-th/9808023, hep-th/9910079, hep-th/0004025;
O.~Lunin and S.~D.~Mathur, hep-th/0006196.

\bibitem{klsc} A.~Klemm and M.~G.~Schmidt,  Phys.\ Lett.\  {\bf B245} (1990) 
53; J.~Fuchs, A.~Klemm and M.~G.~Schmidt, Annals Phys.\  {\bf 214} (1992) 221.

\bibitem{ss} A.~Schwimmer and N.~Seiberg,
Phys.\ Lett.\  {\bf B184} (1987) 191.

\bibitem{mmoores} 
J.~Maldacena, G.~Moore and A.~Strominger,
hep-th/9903163.

\bibitem{gkp} A.~Giveon, D.~Kutasov and O.~Pelc, JHEP {\bf 9910} (1999) 035,
hep-th/9907178.

\bibitem{ags3} R.~Argurio, A.~Giveon and A.~Shomer, 
JHEP {\bf 0004} (2000) 010, hep-th/0002104; 
JHEP {\bf 0012} (2000) 025, hep-th/0011046.

\end{thebibliography}
\end{document}